\documentclass[aps,prd,twocolumn,superscriptaddress,nofootinbib,floatfix]{revtex4-2}

\usepackage{amsmath,amssymb,bm}
\usepackage{graphicx}
\usepackage{hyperref}
\usepackage{xcolor}
\usepackage{booktabs}

\begin{document}

\title{Spinning Billiards and Chaos}

\author{Jacob S.~Lund}
\email{jacob.lund@merton.ox.ac.uk}
\affiliation{Mathematical Institute, University of Oxford, Andrew Wiles Building, Radcliffe Observatory Quarter, Woodstock Road, Oxford, OX2 6GG, U.K.}
\affiliation{The Laboratory for Quantum Gravity \& Strings, Department of Mathematics and Applied Mathematics, University of Cape Town, Cape Town, South Africa}
\affiliation{The National Institute for Theoretical and Computational Sciences, Private Bag X1, Matieland, South Africa}

\author{Jeff Murugan}
\email{jeff.murugan@uct.ac.za}
\affiliation{The Laboratory for Quantum Gravity \& Strings, Department of Mathematics and Applied Mathematics, University of Cape Town, Cape Town, South Africa}
\affiliation{The National Institute for Theoretical and Computational Sciences, Private Bag X1, Matieland, South Africa}

\author{Jonathan P.~Shock}
\email{jonathan.shock@uct.ac.za}
\affiliation{The Laboratory for Quantum Gravity \& Strings, Department of Mathematics and Applied Mathematics, University of Cape Town, Cape Town, South Africa}
\affiliation{The National Institute for Theoretical and Computational Sciences, Private Bag X1, Matieland, South Africa}
\affiliation{Institut National de la Recherche Scientifique, Montreal, Canada}

\date{\today}

\begin{abstract}
We investigate the impact of internal spin on chaos in billiard systems. Extending the standard point-particle billiard by coupling translational and rotational degrees of freedom through a dimensionless spin parameter~$\alpha = I/(mr^2) \in [0,1]$, we find that spin \emph{reduces} chaos monotonically but does \emph{not} eliminate it. In the Bunimovich stadium and Sinai billiard, the Lyapunov exponent decreases with~$\alpha$ but remains positive throughout the physical range, while the circle and rectangle remain integrable. Finite-time Lyapunov exponent distributions reveal a mixed phase space in which spin creates islands of regularity while the majority of trajectories remain chaotic. The mechanism is a conserved quantity $Q = v_\parallel - \alpha u$ preserved through each collision, which constrains the dynamics on sequences of same-orientation wall collisions and explains why spin suppresses chaos more effectively in geometries with longer flat sections. We further show that the Datseris--Hupe--Fleischmann scaling $\lambda \propto 1/f_{\rm chaotic}$ fails for spinning billiards: spin reduces the \emph{intensity} of chaos, not merely the fraction of chaotic trajectories.
\end{abstract}

\maketitle

\section{Introduction}
\label{sec:intro}
\noindent
What happens to chaos when a billiard ball can spin? Classical billiard models, essentially point particles undergoing elastic reflection inside a bounded domain, are paradigmatic systems for studying the interplay of geometry and chaos~\cite{benettin1978,bunimovich1979,sinai1970,chernov2006,tabachnikov2005,gutkin2012}. In large part this is because the geometry of the boundary alone determines the dynamics of the system. Smooth convex tables (such as circles and ellipses) yield integrable motion, while the defocusing mechanism of curved boundaries~\cite{donnay1991,wojtkowski1986} produces positive Lyapunov exponents in the Bunimovich stadium~\cite{bunimovich1979} and Sinai billiard~\cite{sinai1970,benettin1976}.\\

\noindent
Extending these models to particles with internal structure is both physically natural (real balls slip and spin) and mathematically rich. The limiting case of no-slip (perfectly rough) collisions was studied by Broomhead and Gutkin~\cite{broomhead1993}, and has been developed into a systematic framework by Cox, Feres, and collaborators~\cite{cox2017,cox2018,cox2022,cox2021,cox2026}. In the no-slip limit, a rolling constraint couples translation and rotation at each collision, introducing additional conservation laws that can profoundly alter the phase space structure. However, physical collisions are neither perfectly smooth (specular) nor perfectly rough (no-slip); they lie somewhere in between and are controlled by the mass distribution of the ball.\\

\noindent
In this work, we bridge these two limits by introducing a one-parameter family of spinning billiards parametrized by the dimensionless moment of inertia~$\alpha = I/(mr^2) \in [0,1]$. This parameter continuously interpolates between the specular ($\alpha=0$) and maximum-physical-coupling ($\alpha=1$, thin ring) regimes, with the no-slip limit of ~\cite{broomhead1993,cox2026} corresponding to $\alpha \to \infty$.
Through large-scale ensemble computations, we find that spin \emph{reduces} chaos monotonically but does not eliminate it. We analyze the Lyapunov exponent across the full $\alpha$ range (Section~\ref{sec:lyapunov}), examine finite-time Lyapunov exponent distributions to reveal a mixed phase space (Section~\ref{sec:ftle}), and identify a conserved quantity responsible for the chaos reduction (Section~\ref{sec:conserved}). We also investigate dependence on the geometric parameters of the stadium (Section~\ref{sec:geometry_scan}) and the Sinai billiard (Section~\ref{sec:R_scan}).

\section{Setup: Spinning billiard dynamics}
\label{sec:setup}

\subsection{Geometries}
\noindent
While there are many interesting configurations in the literature, for the purposes of our study, we will consider four billiard geometries:
\begin{enumerate}
    \item \textbf{Circle}: A unit disk $x^2 + y^2 \le 1$. This system is integrable for $\alpha=0$.
    \item \textbf{Rectangle}: The 2-dimensional region $[-L,L]\times[-H,H]$ where we will take $L=H=1$ for Lyapunov calculations, and $L=1.5$, $H=1$ for trajectory illustrations. This system is again integrable for $\alpha=0$.
    \item \textbf{Stadium}: The typical Bunimovich stadium consists of flat walls at $y=\pm 1$ for $|x|\le a$ and semicircular caps of unit radius centered at $(\pm a, 0)$, with $a=1$ unless otherwise stated. This system is chaotic for $\alpha=0$.
    \item \textbf{Sinai}: The Sinai billiard consists of the square $[-L,L]^2$ and a central circular obstacle of radius $R$. For our purposes, $L=2$ and $R=1$. This system is also chaotic for $\alpha=0$. This is a finite, hard-walled domain, distinct from the infinite periodic Lorentz gas, although the local scattering dynamics are identical.
\end{enumerate}

\subsection{Collision law}
\noindent
Between collisions, the billiard particle moves freely in straight lines with constant spin~$u = r\omega$, or equivalently the signed surface velocity. Here $r$ is the ball radius and $\omega$ the angular velocity. At each collision with a wall, the normal velocity reverses elastically ($v_{\perp,1} = -v_{\perp,0}$), while the tangential interaction is governed by a restitution parameter $\beta = (1-\alpha)/(1+\alpha)$ that partitions tangential momentum between translational and rotational degrees of freedom consistent with energy conservation~\cite{garwin1969,cross2005}. The resulting update rules are
\begin{align}
    v_{\parallel,1} &= \frac{1-\alpha}{1+\alpha}\, v_{\parallel,0} - \frac{2\alpha}{1+\alpha}\, u_0\,,\label{eq:vpar}\\
    v_{\perp,1} &= -v_{\perp,0}\,,\label{eq:vperp}\\
    u_1 &= -\frac{1-\alpha}{1+\alpha}\, u_0 - \frac{2}{1+\alpha}\, v_{\parallel,0}\,,\label{eq:uspin}
\end{align}
where $v_\parallel = \bm{v}\cdot\hat{\bm{t}}$ and $v_\perp = \bm{v}\cdot\hat{\bm{n}}$ are the velocity components tangent and normal to the boundary at the collision point, with $\hat{\bm{n}}$ the inward-pointing unit normal and $\hat{\bm{t}}$ chosen such that $\hat{\bm{t}}\times\hat{\bm{n}} = +\hat{\bm{z}}$. The parameter $\alpha = I/(mr^2)$ is a dimensionless moment of inertia, made up of the usual the moment of inertia $I$ about the axis perpendicular to the billiard plane through the center of mass,  the mass $m$, and the ball radius, $r$ (the same $r$ appearing in the surface velocity). In what follows, we will set $m = 1$. When $\alpha=0$, these reduce to the usual specular reflection where $v_{\parallel,1} = v_{\parallel,0}$, $v_{\perp,1} = -v_{\perp,0}$, with spin decoupled.\\

\noindent
Equivalently, the collision law~\eqref{eq:vpar} can be characterized by a tangential coefficient of restitution $\beta \equiv v_{\parallel,1}/v_{\parallel,0}|_{u_0=0} = (1-\alpha)/(1+\alpha)$, which decreases monotonically from $\beta = 1$, in the specular case when $\alpha=0$, to $\beta = 0$ when $\alpha=1$. At $\alpha = 1$, the collision law simplifies to $v_{\parallel,1} = -u_0$, $u_1 = -v_{\parallel,0}$ so that tangential velocity and spin are completely exchanged with sign reversal. The no-slip (perfectly rough) limit studied by Cox, Feres, and collaborators~\cite{cox2018,cox2022} corresponds to $\alpha \to \infty$ ($\beta \to -1$); $\alpha = 1$ is the maximum physically realizable value (thin ring), representing zero tangential restitution ($\beta = 0$).\\

\noindent
These equations conserve total kinetic energy (with $m=1$)
\begin{equation}
    E = \tfrac{1}{2} v^2 + \tfrac{1}{2}\alpha u^2 = \text{const}\,,\label{eq:energy}
\end{equation}
which follows from Eqs.~\eqref{eq:vpar}--\eqref{eq:uspin} and can be verified by direct computation.
Measure preservation can also be verified directly. In the $(v_\parallel, u)$ subspace, the collision law acts as a linear map 
\begin{equation}
    M = \begin{pmatrix}
        \frac{1-\alpha}{1+\alpha} & -\frac{2\alpha}{1+\alpha} \\[4pt]
        -\frac{2}{1+\alpha} & -\frac{1-\alpha}{1+\alpha}
    \end{pmatrix},
    \label{eq:jacobian}
\end{equation}
with determinant $\det M = -[(1-\alpha)^2 + 4\alpha]/(1+\alpha)^2 = -1$ so that the map reverses orientation but preserves area. During free flight the spin~$u$ is constant and the spatial dynamics is identical to the spinless case, so the free-flight Jacobian factorizes as the standard Birkhoff factor $\cos\theta'/\cos\theta$ on $(s,\theta)$ times the identity on~$u$. Here $(s,\theta)$ are the standard Birkhoff coordinates where $s$ is the arc-length along the boundary and $\theta$ is the angle of reflection with respect to the inward normal. At collision, $M$~acts on $(v_\parallel, u)$ with $|\det M|=1$, and $v_{\perp,1}=-v_{\perp,0}$. The full collision-to-collision map therefore preserves the extended Liouville measure $\cos\theta\,ds\,d\theta\,du$ on the energy shell.\\

\noindent
The parameter $\alpha = I/(mr^2)$ is the squared ratio of the radius of gyration to the physical radius, and takes different values depending on the mass distribution: $\alpha=2/5$ for a solid sphere, $\alpha=1/2$ for a uniform disk, $\alpha\approx 0.55$ for a tennis ball~\cite{cross2005}, $\alpha=2/3$ for a hollow sphere, and $\alpha=1$ for a thin cylindrical shell. For three-dimensional objects, $\alpha$ is defined with I taken as the moment of inertia about the axis perpendicular to the billiard plane through the center of mass.. Since $I \le mr^2$ for any convex rigid body (with equality when all mass lies on the boundary), the physical range is $\alpha \in [0, 1]$, and all numerical experiments in this work are restricted to this range. Values $\alpha > 1$ are mathematically well-defined but correspond to mass distributions extending beyond the ball surface. We emphasize that the collision law~\eqref{eq:vpar}--\eqref{eq:uspin} assumes a specific tangential restitution model; more complex contact mechanics (e.g., velocity-dependent friction or finite-time contact) could yield different dynamics even at the same~$\alpha$.

\subsection{State space and numerical method}
\noindent
The state of the spinning billiard at each collision is described by five variables: the position $(x,y)$, velocity $(v_x, v_y)$, and spin~$u$. Energy conservation~\eqref{eq:energy} reduces the effective phase space dimensionality by one. The spinless billiard map preserves the Liouville measure $dA = \cos\theta\, ds\, d\theta$ in Birkhoff coordinates and is symplectic on the 2D collision map. With spin, the collision map acts on the 3D space $(s, v_\parallel, u)$ constrained by energy conservation. The dynamics still preserves a natural Liouville measure on this extended phase space, as can be verified by explicit computation of the collision-map Jacobian (see Eq.~\ref{eq:jacobian} below), but the map is volume-preserving rather than symplectic in the strict 2-form sense, since the phase space is odd-dimensional.

\noindent
We compute the leading Lyapunov characteristic number (LCN) via the Benettin renormalization algorithm~\cite{benettin1976,benettin1978}. A reference trajectory and a nearby perturbed trajectory (separated by $\delta_0 = 10^{-7}$ in the full 5D state space) are evolved in parallel. After each collision, the separation is measured and the perturbed trajectory is renormalized back to distance~$\delta_0$ along the separation direction. Perturbations are applied in the full 5D state space rather than projected onto the 4D energy shell. Since energy conservation contributes a zero Lyapunov exponent, the maximal exponent is unaffected by off-shell perturbations, which are rapidly aligned with the most unstable direction by the Benettin renormalization procedure~\cite{benettin1976}. The LCN is computed as the time-averaged logarithmic growth rate of the separation, $\lambda = \lim_{T\to\infty} T^{-1} \sum_i \ln(\delta_i/\delta_0)$, where the sum runs over collisions and $T = \sum_i \Delta t_i$ is the total elapsed time. The Lyapunov exponent is thus defined per unit time (not per collision), which is the natural choice for physical billiards where collision rate varies with $\alpha$ and geometry~\cite{datseris2019}. Ensemble averaging is performed over $10^5$ randomly chosen initial conditions per parameter value, with each trajectory evolved for $10^5$ collisions. This ensemble size was chosen to reduce the standard error of the mean to less than 1\% of the mean value across all $\alpha$ and geometries. Smaller ensembles are used for diagnostics where computational cost is higher or statistical demands are lower: $5{,}000$ ICs for finite time Lyapunov exponent (FTLE) distributions (Section~\ref{sec:ftle}), $8{,}000$ pairs for phase separation (Section~\ref{sec:phase_space}), and $50$ ICs for the full Lyapunov spectrum (Appendix~\ref{app:spectrum}).

\section{Trajectories}
\label{sec:trajectories}
\begin{figure}[t]
    \centering
    \includegraphics[width=\columnwidth]{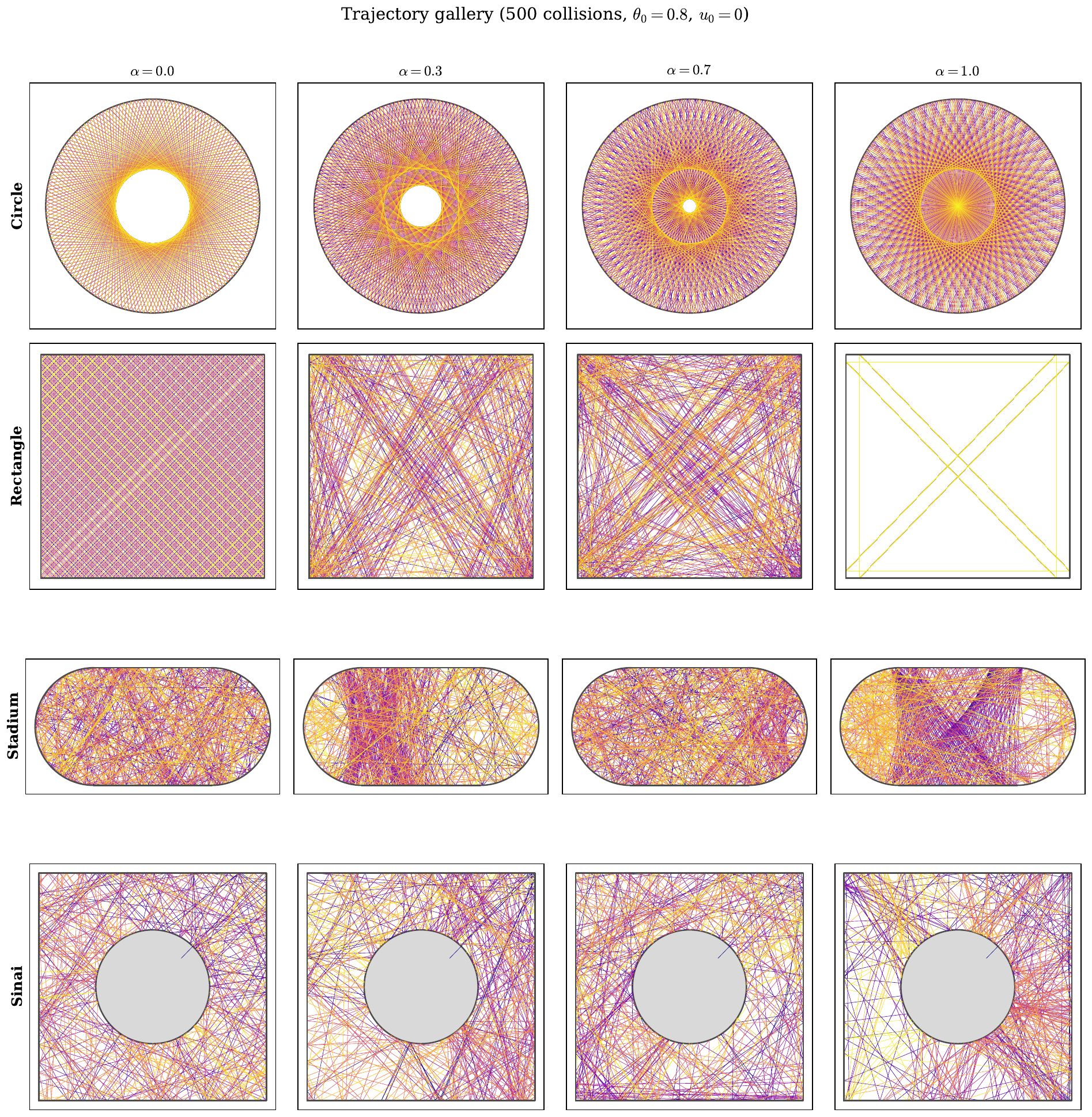}
    \caption{Representative trajectories (500 collisions) for all four geometries at $\alpha = 0$, $0.3$, $0.7$, and $1.0$, colored from early (dark) to late (light) collisions. Spin coupling visibly modifies trajectory structure, creating envelope patterns and apparent regularization, but the stadium and Sinai remain chaotic by the Lyapunov criterion.}
    \label{fig:trajectories}
\end{figure}

\noindent
Figure~\ref{fig:trajectories} shows representative trajectories in all four geometries at several values of~$\alpha$. The qualitative visual differences between specular and spinning billiards are striking---the spin coupling creates envelope structures, caustics, and apparent regularization. However, as our quantitative analysis demonstrates, visual appearance can be misleading;                  the ``regular-looking'' stadium and Sinai trajectories still possess positive Lyapunov exponents.

\section{Lyapunov exponent versus $\alpha$}
\label{sec:lyapunov}
\begin{figure}[t]
    \centering
    \includegraphics[width=\columnwidth]{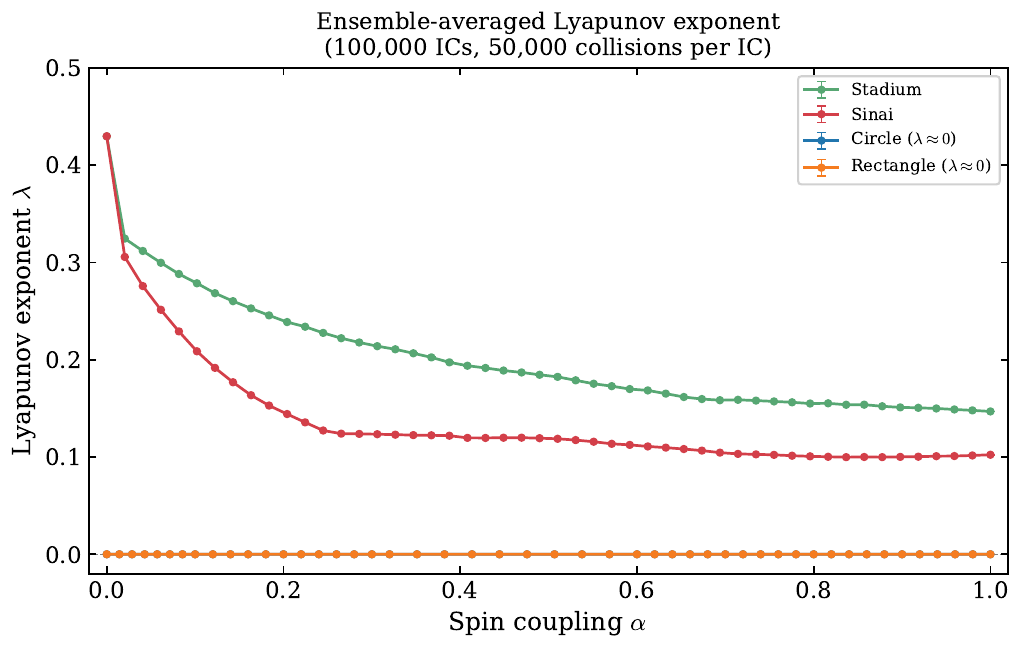}
    \caption{Ensemble-averaged Lyapunov characteristic number as a function of spin parameter~$\alpha$ for all four geometries (50 values of~$\alpha$, $10^5$ ICs, $5\times 10^4$ collisions each for the stadium and Sinai). Error bars show the standard error of the mean. The circle and rectangle remain integrable (LCN~$\approx 0$), while the stadium and Sinai show monotonically decreasing but persistently positive LCN.}
    \label{fig:lyap_sweep}
\end{figure}

\noindent
Figure~\ref{fig:lyap_sweep} shows the ensemble-averaged LCN as a function of~$\alpha$ for all four geometries across 48 values of $\alpha \in [0, 1]$. The key findings are:

\begin{itemize}
    \item \textbf{Circle and Rectangle}: The LCN is consistent with zero ($\lesssim 10^{-4}$) for all~$\alpha$, confirming that the integrable geometries remain integrable under spin coupling.
    \item \textbf{Stadium}: The LCN drops from $\lambda(0) \approx 0.43$ to $\lambda(1) \approx 0.15$---a 66\% reduction---but remains robustly positive.
    \item \textbf{Sinai}: The LCN drops from $\lambda(0) \approx 0.43$ to $\lambda(1) \approx 0.10$---a 76\% reduction---but again remains positive.
\end{itemize}

The $\alpha=0$ values are consistent with published Lyapunov exponents for the standard Bunimovich stadium and Sinai billiard at comparable geometric parameters. Dellago~et~al.~\cite{dellago1996}, for example, found $\lambda \approx 0.37$--$0.44$ depending on geometry, consistent with our value. Our Sinai exponent $\lambda(0) \approx 0.43$ is qualitatively consistent with known results, though direct comparison is complicated by sensitivity to the obstacle-to-box size ratio~\cite{datseris2019}. The positive Lyapunov exponents indicate that the stadium and Sinai billiards remain genuinely chaotic in the presence of spin at all physically realizable values of~$\alpha$. For a solid sphere ($\alpha = 2/5$), the Lyapunov exponent is $\lambda \approx 0.12$ (Sinai) and $\lambda \approx 0.20$ (stadium); even for a thin ring ($\alpha = 1$), chaos persists at $\lambda \approx 0.10$ (Sinai) and $\lambda \approx 0.15$ (stadium).

The two chaotic geometries exhibit qualitatively different dependence on~$\alpha$. The stadium shows a smooth, gradual decay across the full range. The Sinai billiard, by contrast, exhibits two-regime behavior: a rapid initial drop from $\lambda \approx 0.43$ to $\lambda \approx 0.12$ for $\alpha \lesssim 0.3$, followed by a broad plateau at $\lambda \approx 0.10$--$0.12$ for $\alpha \gtrsim 0.3$. This difference has a geometric origin; in the Sinai billiard ($R=1$, $L=2$), only $\sim\!30\%$ of collisions occur with the curved obstacle, creating a clean separation between rapidly-suppressed flat-wall chaos and a persistent curved-wall floor. In the stadium ($a=1$), $\sim\!50\%$ of collisions involve the curved caps, yielding a smoother transition.

\section{Finite-time Lyapunov exponent distributions}
\label{sec:ftle}
\noindent
In standard (spinless) billiards, Birkhoff-coordinate Poincar\'{e} sections $(s, v_\parallel)$ provide a powerful visualization of the phase space, revealing KAM islands, invariant curves, and the chaotic sea directly. In the spinning billiard, however, the collision map acts on the \emph{three}-dimensional space $(s, v_\parallel, u)$ constrained by energy conservation. Projecting onto the two translational coordinates $(s, v_\parallel)$ discards the spin degree of freedom, and no natural two-dimensional invariant surface exists onto which the full dynamics can be reduced. Density-based projections show features at the locations of the well-known bouncing-ball modes, but these cannot be reliably distinguished from projection artifacts without additional analysis.  We therefore characterize the phase space structure using FTLE distributions and chaotic-fraction analysis, which operate on the full $(s, v_\parallel, u)$ phase space without dimensional reduction and provide quantitative measures of the regular-to-chaotic transition.\\

\begin{figure}[t]
    \centering
    \includegraphics[width=\columnwidth]{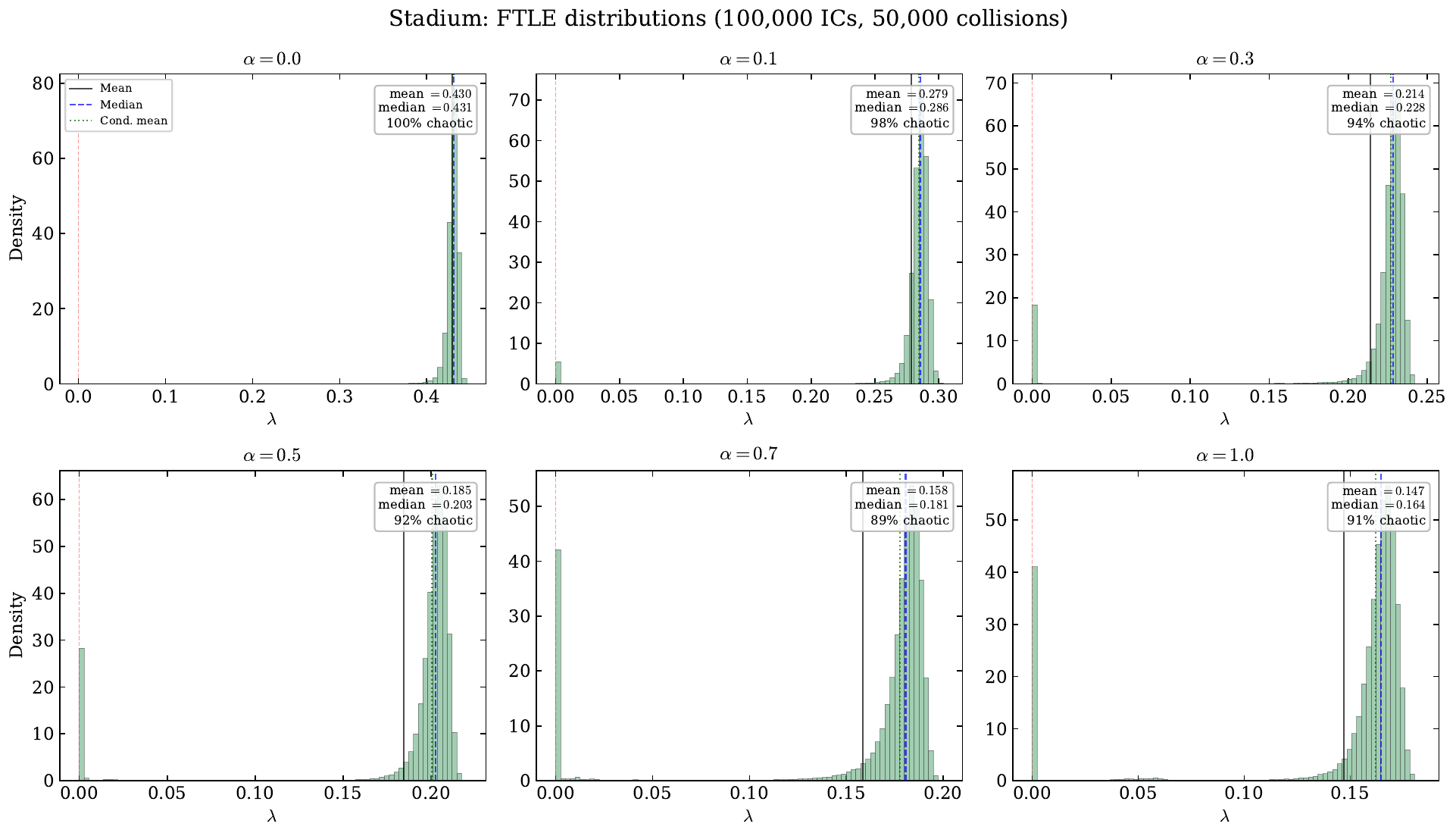}
    \caption{Finite-time Lyapunov exponent distributions for the stadium billiard at six values of~$\alpha$, computed from $10^5$ random initial conditions ($5\times 10^4$ collisions each). Vertical lines show the mean (black, solid), median (blue, dashed), and conditional mean of the chaotic component (green, dotted). The unimodal distribution at $\alpha=0$ becomes bimodal for $\alpha \gtrsim 0.3$, with a secondary peak near zero indicating regular trajectories.}
    \label{fig:ftle}
\end{figure}

\noindent
The mean Lyapunov exponent tells only part of the story. To probe the phase space structure, we compute the FTLE for $10^5$ random initial conditions at each of 6 values of~$\alpha$, with $5\times 10^4$ collisions per trajectory. Figure~\ref{fig:ftle} shows the FTLE distributions for the stadium geometry. At $\alpha=0$, the distribution is tightly unimodal (std~$\approx 0.006$), centered on $\lambda \approx 0.43$---the signature of uniformly chaotic dynamics.\\

\noindent
As $\alpha$ increases, the distributions undergo a qualitative change:
\begin{itemize}
    \item The main peak shifts to lower values and broadens significantly.
    \item A secondary peak emerges near $\lambda\approx 0$, corresponding to trajectories trapped in regular islands.
    \item By $\alpha = 0.5$, the distribution is clearly bimodal: roughly 85--90\% of trajectories remain chaotic (main peak at $\lambda \sim 0.15$--$0.20$), while 10--15\% become regular ($\lambda < 0.05$). Classifying trajectories by their FTLE values and examining the resulting initial-condition distributions, we find that the regular trajectories are predominantly associated with near-bouncing-ball initial conditions: at $\alpha = 0.5$, regular orbits (FTLE~$< 0.01$) have $\sim\!88\%$ of their collisions on the flat walls versus $\sim\!37\%$ for chaotic orbits, and their initial $|v_\parallel|$ is on average $2.4\times$ smaller ($0.27$ vs $0.65$). These are trajectories that bounce nearly perpendicularly between the flat walls, for which long sequences of same-tangent collisions allow $Q$ conservation to constrain the dynamics (see Section~\ref{sec:conserved}).
\end{itemize}

\noindent
This bimodality is the hallmark of a \emph{mixed phase space}: spin creates islands of regular motion embedded within a chaotic sea, rather than globally suppressing chaos. For bimodal distributions, the standard error of the mean is a poor summary statistic as it underestimates the true spread. We therefore also report the median FTLE and the conditional mean restricted to trajectories with positive FTLE (the chaotic component), shown in Fig.~\ref{fig:ftle} as vertical lines. The coexistence of regular islands and a chaotic sea is reminiscent of the ``divided phase space'' in mushroom billiards~\cite{bunimovich2001}, though with a crucial distinction: in mushroom billiards the phase space division is determined by \emph{geometry} (trajectories confined to the cap versus the stem), whereas in spinning billiards the division is \emph{dynamical}, arising from the conserved quantity~$Q$ and the collision-sequence structure. The stickiness of regular-chaotic boundaries~\cite{altmann2005} may also play a role in the broad tails of the FTLE distributions at intermediate~$\alpha$.\\

\noindent
The Sinai billiard exhibits qualitatively similar behavior. Notably, the coexistence of regular islands and a chaotic sea implies that spinning billiards with $\alpha > 0$ are \emph{not} ergodic, in contrast to the $\alpha = 0$ case where the Bunimovich stadium~\cite{bunimovich1979} and Sinai billiard~\cite{sinai1970} are proven ergodic~\cite{chernov2006}. The breakdown of ergodicity represents a qualitative change in the dynamical character induced by spin coupling.

\section{Chaotic fraction}
\label{sec:chaotic_fraction}
\begin{figure}[t]
    \centering
    \includegraphics[width=\columnwidth]{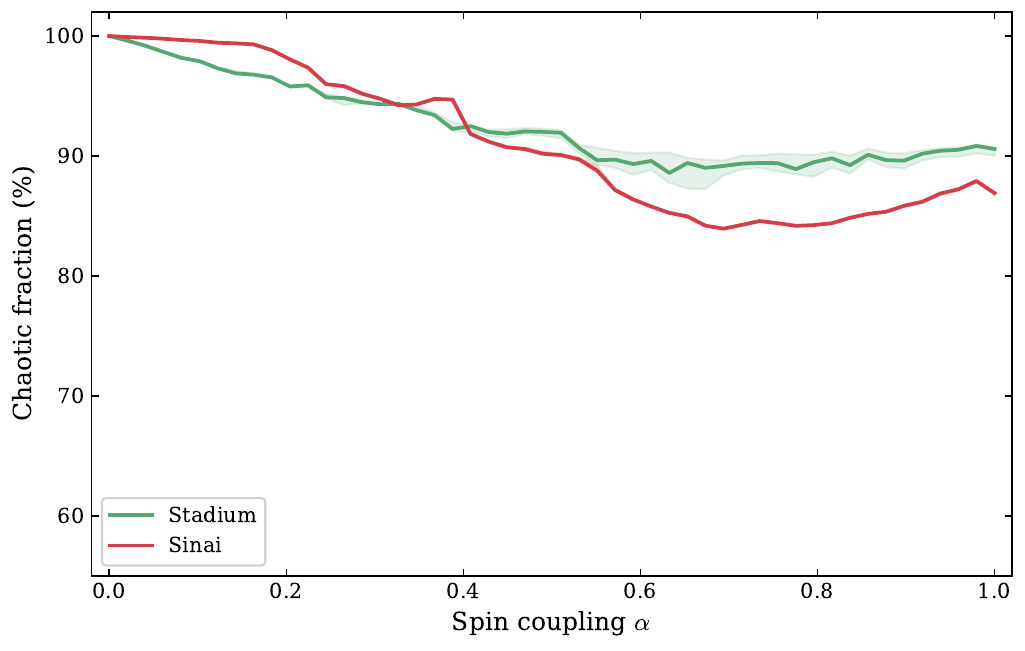}
    \caption{Chaotic fraction (percentage of initial conditions with FTLE $> 0.01$) as a function of~$\alpha$ for the stadium and Sinai geometries ($10^5$ ICs, $5\times 10^4$ collisions each, 50 values of~$\alpha$). The shaded bands show the range obtained by varying the FTLE threshold between $0.001$ and $0.05$; the near-invisibility of the bands confirms robustness to the threshold choice.}
    \label{fig:chaotic_fraction}
\end{figure}

\noindent
We quantify the fraction of phase space that remains chaotic by classifying each trajectory as ``chaotic'' (FTLE $> 0.01$) or ``regular'' (FTLE $\le 0.01$). Using $10^5$ initial conditions per $\alpha$ value across 50 values of $\alpha \in [0, 1]$, we compute the chaotic fraction $f_{\rm ch}(\alpha)$. The results are robust to the choice of threshold; varying the FTLE cutoff between $0.001$ and $0.05$ shifts the chaotic fraction by less than 1~percentage point (shown as shaded bands in Fig.~\ref{fig:chaotic_fraction}).
Figure~\ref{fig:chaotic_fraction} shows the results. Some key observations that follow from this figure are:
\begin{itemize}
    \item At $\alpha=0$, 100\% of trajectories are chaotic in both the stadium and Sinai.
    \item The chaotic fraction decreases gradually with~$\alpha$, reaching $\sim\!90\%$ for the stadium and $\sim\!87\%$ for the Sinai at $\alpha = 1$. (A slight uptick in the Sinai curve near $\alpha \approx 1$ is within statistical uncertainty and does not appear consistently across threshold choices.)
    \item Even for a thin ring ($\alpha = 1$), the vast majority of initial conditions produce chaotic trajectories, confirming that spin creates isolated regular islands rather than globally regularizing the dynamics.
\end{itemize}

\section{Phase space analysis}
\label{sec:phase_space}

\subsection{Phase space separation}
\noindent
To complement the Lyapunov analysis, we study the growth of separation between nearby trajectories in phase space. Starting from 8000 pairs of initial conditions separated by $\delta_0 = 10^{-7}$ in a random direction in the 5D state space, we compute $\ln(\delta_n/\delta_0)$ as a function of collision number~$n$.\\

\begin{figure}[t]
    \centering
    \includegraphics[width=\columnwidth]{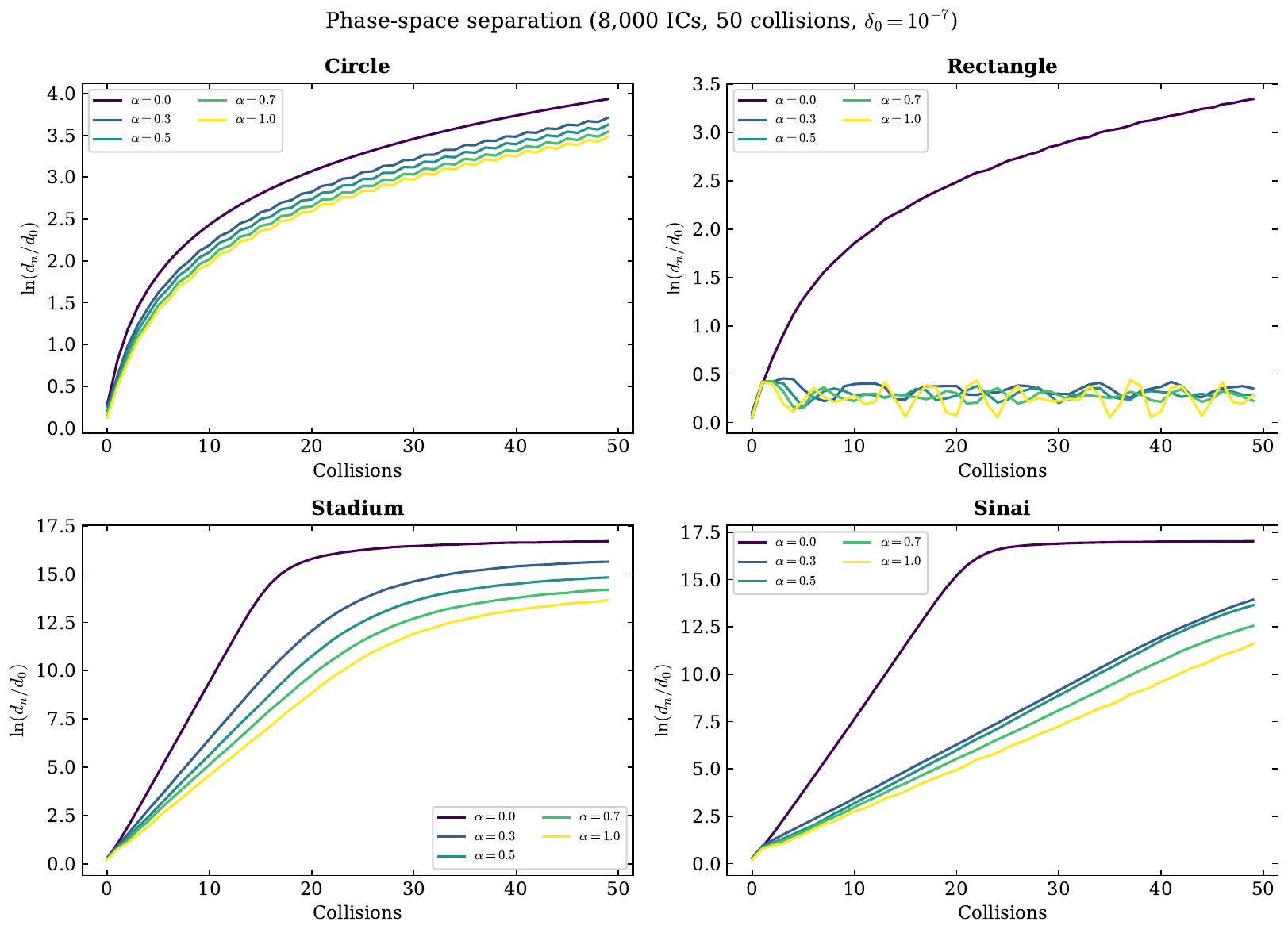}
    \caption{Phase space separation $\ln(\delta_n/\delta_0)$ vs collision number for the four geometries at several $\alpha$ values ($8{,}000$ trajectory pairs, $\delta_0 = 10^{-7}$, 50 collisions). Linear growth (stadium, Sinai) indicates exponential divergence and genuine chaos; logarithmic growth (circle, rectangle) indicates integrable dynamics.}
    \label{fig:separation}
\end{figure}

\noindent
Figure~\ref{fig:separation} shows the results for all four geometries at representative $\alpha$ values over 50 collisions. (The separation remains well below the system size throughout this window: at $\lambda \approx 0.3$ and $\delta_0 = 10^{-7}$, the separation after 50 collisions is $\sim\! 10^{-7} e^{15} \approx 10^{-1}$, far from the billiard diameter.) For the circle and rectangle, $\ln(\delta_n/\delta_0)$ grows logarithmically (sublinearly), consistent with integrable dynamics. For the stadium and Sinai, $\ln(\delta_n/\delta_0)$ grows \emph{linearly} at all~$\alpha$ values, confirming exponential divergence $\delta_n \sim \delta_0\,e^{\lambda n}$---the defining signature of chaos. The slope decreases with~$\alpha$, matching the Lyapunov exponent results. (The quantitative growth rate is measured by the Benettin algorithm over $2\times 10^5$ collisions in Section~\ref{sec:lyapunov}; this short-time window serves to confirm the divergence type---exponential versus polynomial.)

\section{Conserved quantity and mechanism}
\label{sec:conserved}
\noindent
The linear combination
\begin{equation}
    Q \equiv v_\parallel - \alpha\, u\,,
    \label{eq:Q}
\end{equation}
is conserved under the collision law~\eqref{eq:vpar}--\eqref{eq:uspin}. To see this, note that
\begin{align}
    Q_1 &= v_{\parallel,1} - \alpha\, u_1 \notag\\
    &= \frac{1-\alpha}{1+\alpha}v_{\parallel,0} - \frac{2\alpha}{1+\alpha}u_0 + \alpha\left(\frac{1-\alpha}{1+\alpha}u_0 + \frac{2}{1+\alpha}v_{\parallel,0}\right)\notag\\
    &= v_{\parallel,0} - \alpha\, u_0 = Q_0\,.
    \label{eq:Q_conservation}
\end{align}
Thus $Q$ is exactly conserved through \emph{each individual collision}, regardless of whether the wall is flat or curved. This is a purely algebraic consequence of the collision law and holds for all wall types. However, $Q$ does  depend on the tangential velocity $v_\parallel = \bm{v}\cdot\hat{\bm{t}}$, which is defined relative to the \emph{local} tangent direction~$\hat{\bm{t}}$ at each collision point. Between collisions, the velocity~$\bm{v}$ and spin~$u$ are constant (free flight), but when the particle hits a wall segment with a different tangent direction, the decomposition of~$\bm{v}$ into tangential and normal components changes. Consequently:

\begin{itemize}
    \item On a sequence of collisions with wall segments sharing the \emph{same} tangent direction, the value of~$Q$ is exactly preserved (to machine precision, $\sim 10^{-16}$).
    \item At a \emph{transition} between wall segments with different tangent directions (e.g., flat wall $\to$ curved cap, or top wall $\to$ right wall), $Q$ exhibits an $\mathcal{O}(1)$ jump because $v_\parallel$ is measured relative to different tangent vectors.
\end{itemize}
\noindent
To understand why $Q$-conservation reduces chaos when $\alpha > 0$ but not when $\alpha = 0$, consider the effective dimensionality of the dynamics. At $\alpha = 0$ (no spin), the state space is the four dimensional space of $(x, y, v_x, v_y)$. Energy conservation reduces it to three dimensions. The collision map in Birkhoff coordinates $(s, v_\parallel)$ is 2-dimensional; $Q = v_\parallel$ is trivially conserved at each specular reflection, but this provides no constraint beyond what is already encoded in the Birkhoff map; $v_\parallel$ is simply one of its two coordinates.\\

\noindent
When $\alpha > 0$, the state space is now five dimensional with coordinates $(x, y, v_x, v_y, u)$. Energy conservation reduces it by one dimension, and at each collision the Birkhoff map is 3-dimensional with coordinates $(s, v_\parallel, u)$. Now $Q = v_\parallel - \alpha u$ provides a \emph{genuinely new} constraint that couples the spin and translational degrees of freedom. On sequences of same-tangent collisions, $Q$ conservation reduces the 3D Birkhoff map back to 2D, the same effective dimensionality as the spinless billiard. The net effect is that spin introduces both a new degree of freedom and a new conservation law; on same-tangent collision sequences, these offset one another, reducing the dynamics to a 2D map equivalent in dimensionality to the standard spinless Birkhoff map. The constrained 2D map is not identical to the spinless Birkhoff map since the relationship $v_\parallel = Q + \alpha u$ couples the translational and rotational variables, but the dimensional reduction removes the additional expanding direction that would otherwise be available. At wall transitions where the tangent direction changes, $Q$ undergoes $\mathcal{O}(1)$ jumps and the full 3D dynamics is briefly restored, maintaining chaos. (Note that the integrability of the spinning circle billiard follows from its continuous rotational symmetry, not from $Q$ conservation, since the tangent direction changes at every collision on a curved wall. More generally, $Q$ conservation alone, together with energy, does not suffice for integrability in any geometry: it provides only one additional constraint on the 3D collision map, leaving the dynamics 2D rather than 1D.) We verify the conservation of $Q$ numerically in Appendix~\ref{app:conserved_quantity}. Tracking $|\Delta Q| = |Q_{n+1} - Q_n|$ across $10{,}000$ collisions in the stadium reveals two sharply separated populations: same-wall consecutive pairs at machine precision ($\sim 10^{-16}$) and wall-transition pairs at $\mathcal{O}(1)$---a gap of ${\sim}\,16$ orders of magnitude.\\

\noindent
This provides a clean mechanistic explanation for why spin reduces chaos. On sequences of collisions with \emph{flat} walls, where consecutive tangent directions are identical, $Q$~conservation (together with energy) constrains the 3D Birkhoff map $(s, v_\parallel, u)$ to a 2D surface. Since $Q = v_\parallel - \alpha u$ is fixed, the spin is slaved to translation: $u = (v_\parallel - Q)/\alpha$. Together with the normal-velocity reversal $v_{\perp,1} = -v_{\perp,0}$, this yields an effectively \emph{integrable} map, equivalent to specular reflection on a flat wall carrying an additional conserved phase-space label~$Q$. The Lyapunov exponent on these flat-wall sequences is exactly zero. On curved walls, by contrast, the tangent frame rotates between collisions, so the value of~$Q$ as measured in successive tangent frames generically differs, even though $Q$ is exactly conserved through each individual collision; consequently, the integrable reduction does not apply. This is precisely why curved-wall geometries maintain chaos. At wall transitions where the tangent direction changes discretely (e.g., flat wall to curved cap), $Q$ undergoes $\mathcal{O}(1)$ jumps and the full 3D dynamics is restored, producing positive local stretching rates. The observed $\lambda(\alpha)$ is thus a weighted average of zero contributions (flat-wall, integrable sequences) and positive contributions (wall-transition events), with the weights shifting toward the integrable component as~$\alpha$ increases.\\

\noindent
We can sharpen this further by measuring the fraction of wall-transition collisions~$f_{\rm trans}(\alpha)$ directly. In the stadium, $f_{\rm trans}$ is nearly constant at $\sim\!80\text{--}84\%$ across all~$\alpha$ ($1{,}000$ ICs, $10{,}000$ collisions). In the Sinai billiard, $f_{\rm trans} = 1$ identically, since the circular obstacle ensures that every collision changes the tangent direction. Since~$f_{\rm trans}$ is approximately $\alpha$-independent in both geometries, the entire $\lambda(\alpha)$ decrease must originate from a reduction in the \emph{local stretching rate at wall transitions}, not from a shift in the fraction of transition events.

\section{Dependence on stadium geometry}
\label{sec:geometry_scan}
\noindent
The stadium geometry is parameterized by the half-length~$a$ of the flat section. At $a=0$ the stadium reduces to a circle (integrable), and for $a>0$ it is generically chaotic. We investigate how the $\lambda(\alpha)$ curve depends on~$a$.\\

\begin{figure}[t]
    \centering
    \includegraphics[width=\columnwidth]{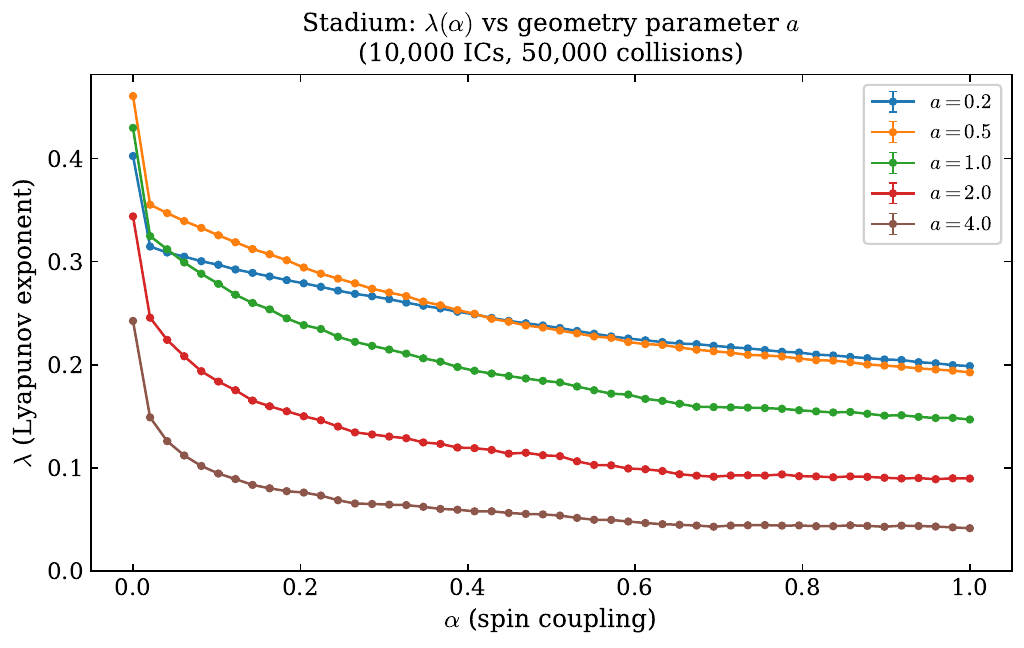}
    \caption{Lyapunov exponent vs~$\alpha$ for stadiums with different flat-section lengths~$a$ ($10{,}000$ ICs, $5\times 10^4$ collisions each). Longer flat sections lead to greater chaos reduction, consistent with the conserved-quantity mechanism.}
    \label{fig:geometry}
\end{figure}

\noindent
Figure~\ref{fig:geometry} shows $\lambda(\alpha)$ for five values of $a \in \{0.2, 0.5, 1.0, 2.0, 4.0\}$. The initial Lyapunov exponent $\lambda(0)$ increases with~$a$, and spin reduces it by a larger absolute amount for stadiums with longer flat sections. However, the \emph{relative} reduction $\lambda(\alpha)/\lambda(0)$ is roughly $a$-independent, suggesting a universal scaling behavior. The physical interpretation is clear: longer flat sections mean a higher fraction of collisions where consecutive wall tangents are aligned, giving the $Q$-conservation mechanism more leverage. But even at $a=4.0$, the LCN remains positive for all~$\alpha$, indicating that the curved caps, however rarely visited, maintain the chaotic mixing.\\

\begin{figure}[t]
    \centering
    \includegraphics[width=\columnwidth]{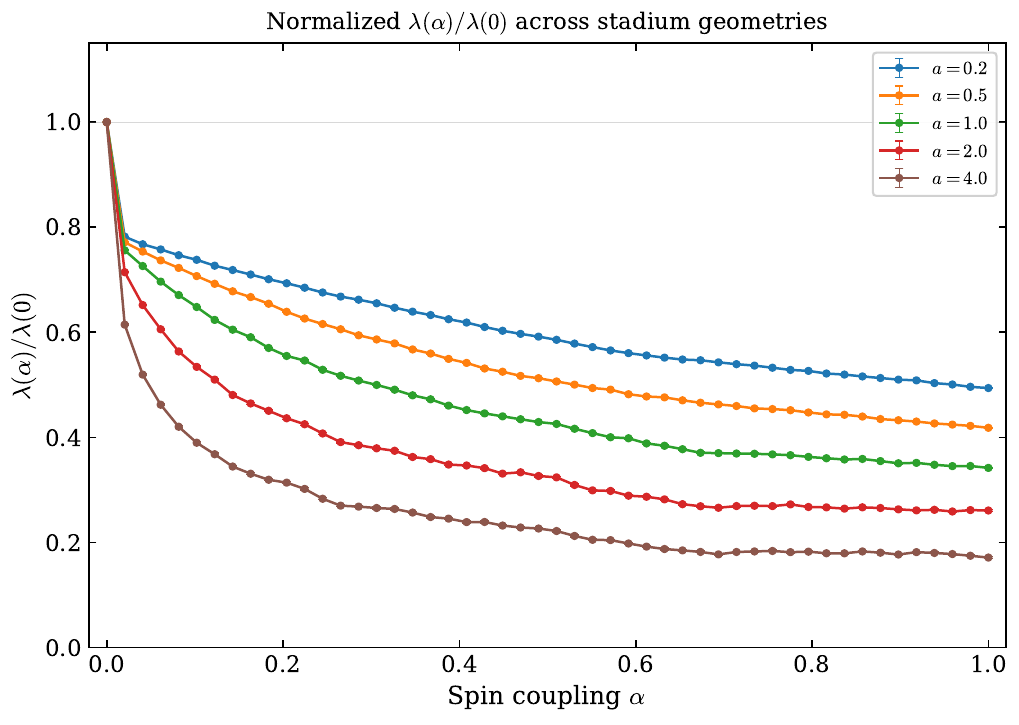}
    \caption{Normalized ratio $\lambda(\alpha)/\lambda(0)$ for stadiums with five different flat-section lengths~$a$. The curves do not collapse: longer flat sections produce greater relative chaos reduction, consistent with the $Q$-conservation mechanism having more leverage when a higher fraction of collisions share the same tangent direction. All curves remain bounded away from zero, confirming persistent chaos at all~$\alpha$.}
    \label{fig:collapse}
\end{figure}

\noindent
To test whether the relative chaos reduction is geometry-independent, Fig.~\ref{fig:collapse} shows the normalized ratio $\lambda(\alpha)/\lambda(0)$ for all five stadium geometries. The curves do \emph{not} collapse: longer flat sections (larger~$a$) produce systematically greater relative chaos reduction. This is consistent with the $Q$-conservation mechanism having more leverage when a larger fraction of collisions share the same tangent direction. All curves remain bounded away from zero, confirming persistent chaos at every~$\alpha$.

\section{Obstacle radius dependence in the Sinai billiard}
\label{sec:R_scan}

\begin{figure}[t]
    \centering
    \includegraphics[width=\columnwidth]{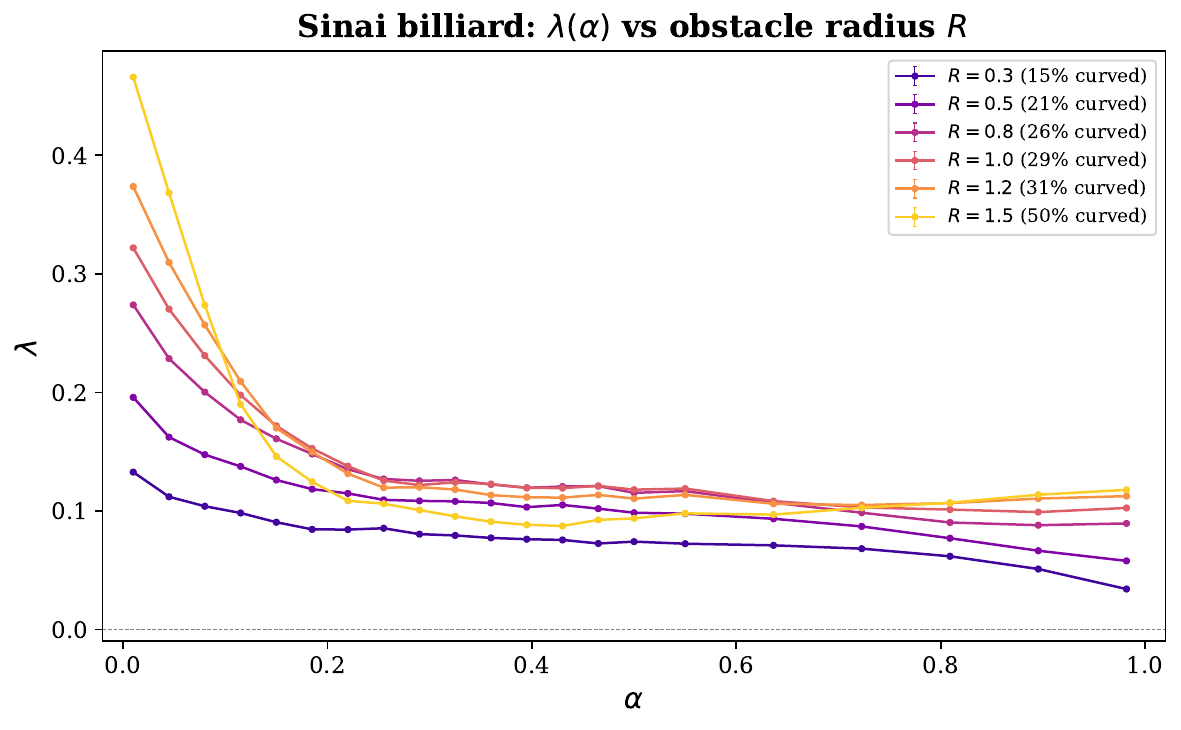}
    \caption{Lyapunov exponent vs~$\alpha$ for the Sinai billiard with different obstacle radii~$R$ (square side $L=2$), computed with 500 initial conditions per point with error bars showing SEM. Parenthetical percentages show the fraction of collisions hitting the curved obstacle. All curves show a fast initial drop followed by a plateau whose height increases with~$R$.}
    \label{fig:R_scan}
\end{figure}

To further test the two-regime mechanism, we vary the obstacle radius~$R$ in the Sinai billiard ($L=2$) and compute $\lambda(\alpha)$ for $\alpha \in [0, 1]$, using 500 initial conditions per point. As $R$ increases, the fraction of collisions with the curved obstacle grows from $\sim\!15\%$ ($R=0.3$) to $\sim\!50\%$ ($R=1.5$). Figure~\ref{fig:R_scan} shows the results. The two-regime structure, a fast drop followed by plateau, is visible for all~$R$, with several noteworthy features:
\begin{itemize}
    \item For small~$R$ (e.g.\ $R=0.3$, 15\% curved): the initial Lyapunov exponent is lower ($\lambda(0) \approx 0.13$), and the plateau settles at $\lambda \approx 0.07$.
    \item For intermediate~$R$ ($R=0.8$--$1.0$, 26--29\% curved): a clear fast drop is followed by a well-defined plateau at $\lambda \approx 0.10$--$0.12$.
    \item For large~$R$ ($R=1.5$, 50\% curved): the initial exponent is much higher ($\lambda(0)\approx 0.47$), and the Lyapunov exponent exhibits a \emph{non-monotonic} feature: after the fast initial drop, $\lambda$ partially recovers around $\alpha \approx 0.5$--$1.0$ before reaching the plateau. The origin of this non-monotonic behavior is not fully understood; it may reflect a resonance between constrained flat-wall dynamics and curved-wall scattering that emerges when curved collisions constitute a large fraction of the total, and warrants further investigation. The feature is robust to ensemble size (persisting from 250 to 1000~ICs) and is also present in the per-collision Lyapunov exponent (Appendix~\ref{app:collision_rate}), ruling out both finite-sample artifacts and collision-rate effects.
    \item The plateau height $\lambda_\infty$ increases with~$R$, consistent with the physical picture: a larger curved obstacle produces more curved-wall collisions, sustaining a higher level of irreducible chaos.
\end{itemize}

\section{Discussion and conclusions}
\label{sec:conclusions}
\noindent
Our analysis reveals a nuanced picture of how internal spin modifies chaos in billiard systems. Some key results are:

\begin{enumerate}
    \item \textbf{Spin reduces but does not eliminate chaos.} The Lyapunov exponent decreases monotonically with~$\alpha$ for the stadium and Sinai billiards, but remains robustly positive throughout the physical range $\alpha \in [0, 1]$. Even for a thin ring ($\alpha=1$, the maximum for any rigid body), the Sinai Lyapunov exponent retains $\sim\!23\%$ of its specular value.

    \item \textbf{Geometry-dependent behavior.} The two chaotic geometries exhibit qualitatively different dependence on~$\alpha$: the stadium shows a smooth, gradual decay, while the Sinai billiard shows a rapid initial drop followed by a persistent plateau at $\lambda \approx 0.10$. Stadiums with longer flat sections and Sinai billiards with smaller obstacles show greater chaos reduction, consistent with the proportion of ``constrained'' (same-tangent) versus ``unconstrained'' (tangent-changing) collision sequences.

    \item \textbf{A mixed phase space.} FTLE distributions transition from unimodal (uniformly chaotic) at $\alpha=0$ to bimodal (mixed regular/chaotic) for $\alpha\gtrsim 0.3$, with $\sim\!85\%$ of trajectories remaining chaotic at $\alpha=1$.

    \item \textbf{A clear mechanism.} The conserved quantity $Q=v_\parallel - \alpha u$ is preserved through each individual collision (flat and curved alike), but its value---measured relative to the local wall tangent---jumps at transitions between differently-oriented wall segments. This provides an additional constraint on sequences of same-orientation collisions, reducing the effective dimensionality of the dynamics.
\end{enumerate}
\noindent
Our findings connect naturally to the growing literature on no-slip billiards~\cite{broomhead1993,cox2017,cox2018,cox2022,cox2021,cox2026}. The conserved quantity~$Q = v_\parallel - \alpha u$ is the finite-$\alpha$ generalization of the rolling-velocity projection that plays a central role in the no-slip billiard theory of Cox and Feres~\cite{cox2017,cox2021}; the novelty here is that it remains exactly conserved at all~$\alpha$ and provides a sharp explanation for chaos reduction across the full physical range. The regularizing effect of $Q$ conservation aligns with the observation by Cox and Feres that constructing chaotic no-slip billiards is nontrivial~\cite{cox2022}. However, our results show that for curved boundaries, the spin-velocity coupling is insufficient to eliminate chaos within the physical range of~$\alpha$, and the persistently positive Lyapunov exponent suggests that curved-wall chaos is a robust feature of these systems.\\

\noindent
Two additional quantitative tests strengthen this picture. First, the Datseris--Hupe--Fleischmann scaling $\lambda \propto 1/f_{\rm chaotic}$~\cite{datseris2019} fails for spinning billiards (see Appendix~\ref{app:dh_scaling}): spin reduces the intensity of chaos within the chaotic component, not merely the size of the chaotic sea. This distinguishes spin coupling from perturbations that create regular islands while leaving the chaotic dynamics unchanged---here, the instability of chaotic trajectories is itself suppressed by~$Q$ conservation at wall transitions. Second, the full Lyapunov spectrum (Appendix~\ref{app:spectrum}) reveals that only one exponent is positive at all~$\alpha$---spin coupling does not create hyperchaos, consistent with the dimensional-reduction mechanism of~$Q$ conservation.\\

\noindent
The no-slip billiards of Cox and Feres~\cite{cox2017,cox2021,cox2026} correspond to the limit $\alpha \to \infty$ ($\beta \to -1$), which lies outside the physical range $\alpha \in [0,1]$ for any rigid body (since $\alpha = I/mr^2 \leq 1$). Exploring this mathematical extension could reveal whether the monotonic decrease in~$\lambda$ continues or whether qualitatively new behavior emerges at $\beta < 0$. The broader implication is that internal degrees of freedom can significantly modify but not necessarily destroy the chaotic character of a dynamical system, which would be relevant to granular flows, soft matter, and other contexts where extended particles interact with confining geometries.\\

\noindent
Future directions and extensions of this work include: (i)~extending to three-dimensional spin (vector angular velocity), (ii)~understanding the non-monotonic feature observed for large Sinai obstacle radii ($R=1.5$), (iii)~exploring the mathematical extension $\alpha > 1$ to connect with the no-slip limit, (iv)~connections to quantum billiard systems where internal spin degrees of freedom may affect level statistics, (v)~characterizing the topology and connectivity of the regular islands in the extended phase space $(s, v_\parallel, u)$, including how they grow, merge, or fragment as~$\alpha$ increases, and whether their boundaries exhibit the stickiness phenomena observed in mushroom billiards~\cite{altmann2005}, and (vi)~analytical (linearized) arguments for the persistence of positive Lyapunov exponents on curved boundaries, complementing the present numerical evidence.

\section*{Data and code availability}
The code used to generate all figures in this paper is available at \href{https://github.com/shocklab/Spinning-billiards}{https://github.com/shocklab/Spinning-billiards}

\section*{Acknowledgments}
We are especially grateful to Francois Kemp for his collaboration in the early stages of this work, and to Haris Skokos and San\'e Erasmus for their helpful suggestions. JM would like to acknowledge support from the ICTP through the Associates Programme, from the Simons Foundation (Grant No.\ 284558FY19), and from the ``Quantum Technologies for Sustainable Development'' grant from the National Institute for Theoretical and Computational Sciences of South Africa (NITHECS). JS would like to acknowledge support from the Artificial Intelligence for Development (AI4D) program, a partnership between IDRC and the UK's Foreign, Commonwealth and Development Office. Claude Code assisted with code optimisation, experimental design, theoretical development, and framing.

\appendix
\section{Numerical verification}
\label{app:verification}

\subsection{Convergence of LCN estimates}
\begin{figure}[t]
    \centering
    \includegraphics[width=\columnwidth]{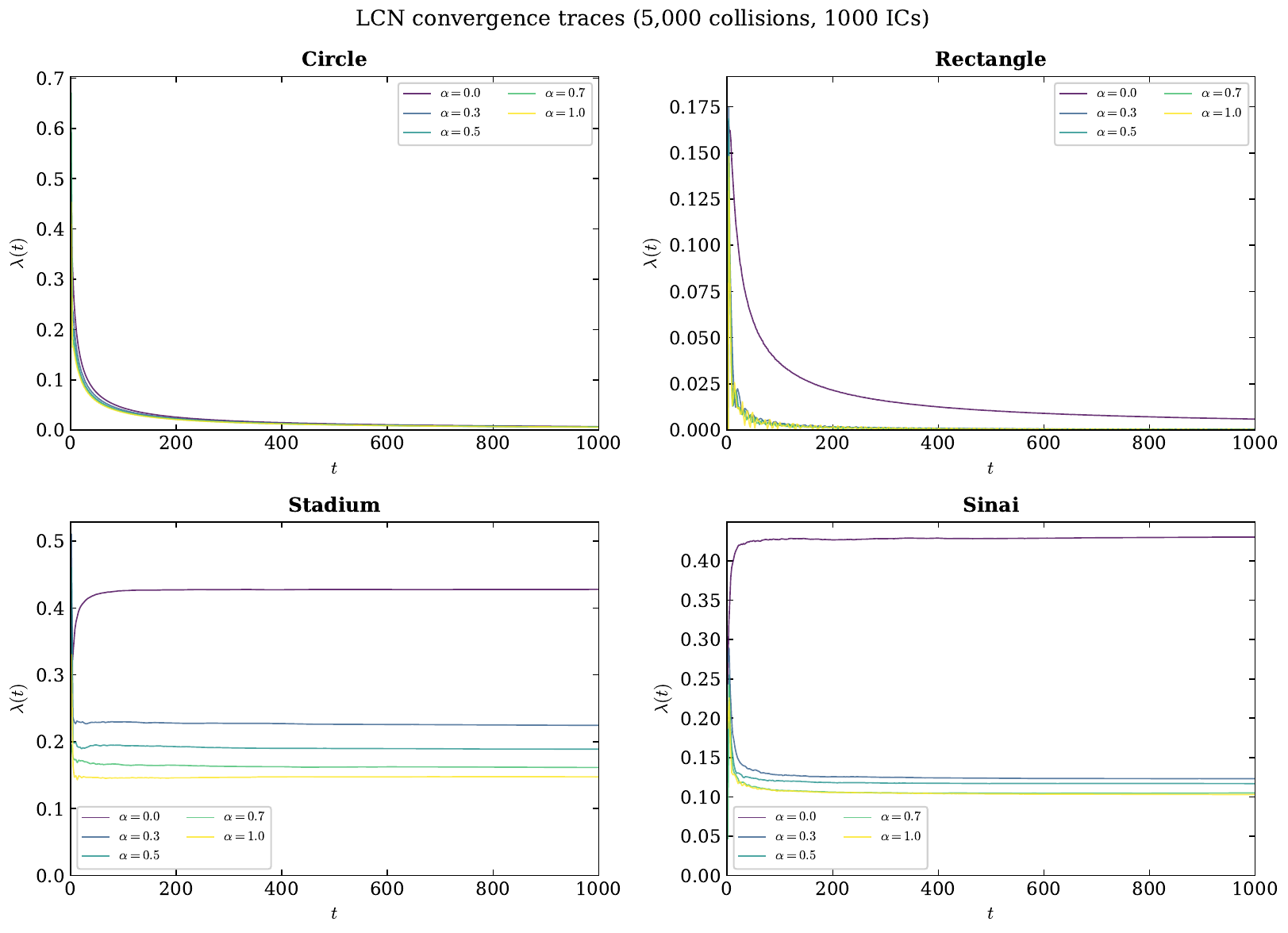}
    \caption{Single-trajectory LCN convergence traces for all four geometries at five values of~$\alpha$ ($10^5$ collisions, $\theta_0=0.8$, $u_0=0$). The circle and rectangle show $\sim 1/t$ decay to zero. The stadium and Sinai converge to positive plateaus, ruling out ``slow convergence to zero.''}
    \label{fig:lcn_traces}
\end{figure}

\noindent
To ensure that the finite-time LCN estimates are well-converged, we examine convergence from two perspectives. Figure~\ref{fig:lcn_traces} shows single-trajectory LCN curves $\lambda(t)$ as a function of cumulative time for all four geometries at several $\alpha$ values. For the circle and rectangle, the LCN decays as $\sim 1/t$ (power-law), consistent with zero limiting value. For the stadium and Sinai, the LCN curves level off to clear plateaus by $\sim 10^4$ collisions, confirming that our $5\times 10^4$-step estimates are converged. Table~\ref{tab:convergence} shows the ensemble-averaged LCN as a function of the number of collisions used in each trajectory ($10{,}000$ ICs per point). The estimates are stable to three significant figures across all collision counts, confirming convergence well before our default of $5\times10^4$ collisions.

\begin{table}[t]
    \caption{Ensemble-averaged LCN ($\pm$ SEM) vs.\ number of collisions for the stadium and Sinai billiards ($10{,}000$ ICs per entry). Numbers in parentheses denote the standard error of the mean in the last displayed digit.}
    \label{tab:convergence}
    \begin{ruledtabular}
    \begin{tabular}{llcccc}
    & & \multicolumn{4}{c}{Number of collisions} \\
    \cline{3-6}
    Geometry & $\alpha$ & $10^4$ & $5\!\times\!10^4$ & $10^5$ & $5\!\times\!10^5$ \\
    \hline
    Stadium & 0.1 & 0.281(0) & 0.280(0) & 0.279(0) & 0.280(0) \\
            & 0.5 & 0.183(1) & 0.183(1) & 0.183(1) & 0.183(1) \\
            & 1.0 & 0.147(1) & 0.147(1) & 0.146(1) & 0.147(1) \\
    \hline
    Sinai   & 0.1 & 0.211(0) & 0.211(0) & 0.210(0) & 0.210(0) \\
            & 0.5 & 0.120(0) & 0.119(0) & 0.119(0) & 0.119(0) \\
            & 1.0 & 0.103(0) & 0.103(0) & 0.102(0) & 0.102(0) \\
    \end{tabular}
    \end{ruledtabular}
\end{table}

\subsection{Energy conservation}

\begin{figure}[t]
    \centering
    \includegraphics[width=\columnwidth]{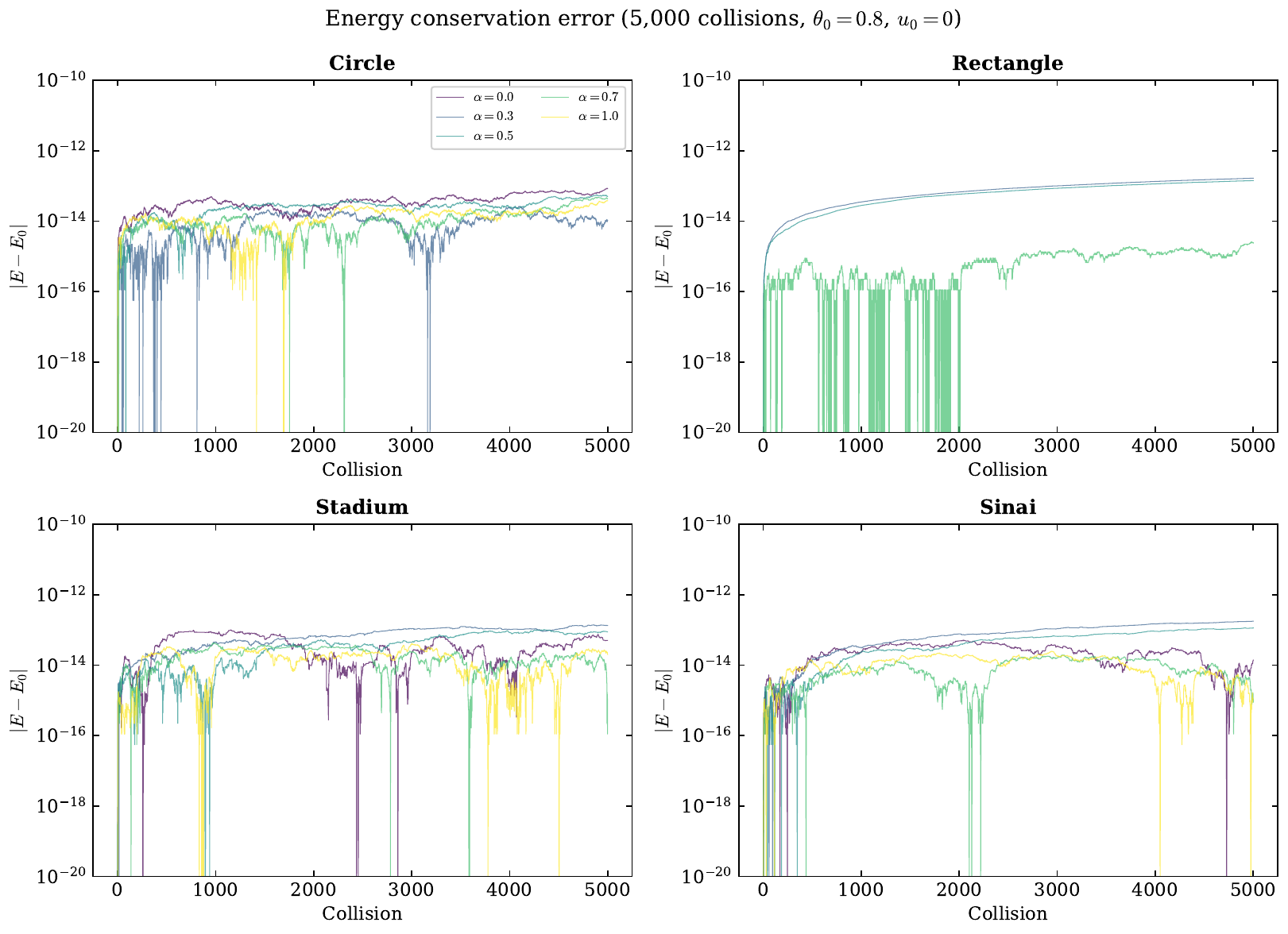}
    \caption{Energy deviation $|E_n - E_0|$ vs collision number for all four geometries at several~$\alpha$ values ($5{,}000$ collisions, $\theta_0=0.8$, $u_0=0$). Energy is conserved to machine precision ($\sim 10^{-15}$) throughout.}
    \label{fig:energy}
\end{figure}

As a numerical diagnostic, we verify that the total kinetic energy $E = \frac{1}{2}(v_x^2 + v_y^2 + \alpha u^2)$ is conserved throughout each simulation. Figure~\ref{fig:energy} shows the absolute energy deviation $|E_n - E_0|$ as a function of collision number. Energy is conserved to machine precision ($\sim 10^{-15}$) in all geometries and at all values of~$\alpha$, confirming the correctness of the collision law implementation.

\subsection{Collision rate}
\label{app:collision_rate}
\begin{figure}[t]
    \centering
    \includegraphics[width=\columnwidth]{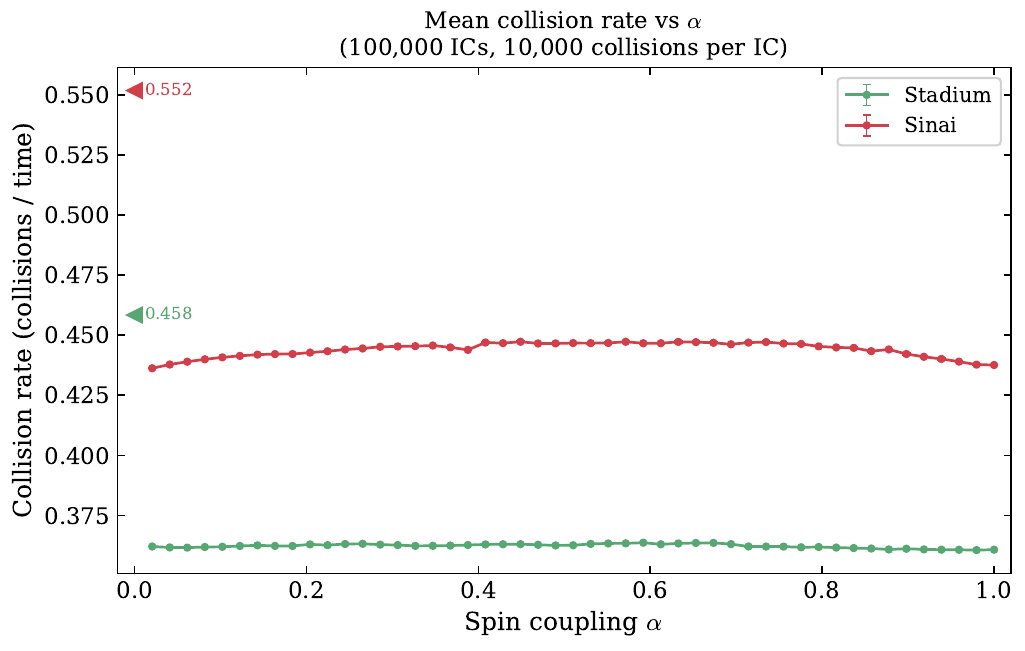}
    \caption{Mean collision rate vs~$\alpha$ for the stadium and Sinai geometries ($10^5$ ICs, $10{,}000$ collisions per IC). At $\alpha=0$ (no spin), all energy is translational, giving a higher collision rate (triangles on the $y$-axis). For $\alpha > 0$ the rate is essentially flat: ${\lesssim}\,1\%$ variation for the stadium and ${\lesssim}\,3\%$ for Sinai, far smaller than the $\sim\!70\%$ change in the Lyapunov exponent, confirming that the $\lambda(\alpha)$ trends are not an artifact of collision-rate changes. The visual gap between $\alpha=0$ and the smallest $\alpha > 0$ reflects the discrete sampling; the rate approaches the $\alpha=0$ value continuously as $\alpha\to 0^+$.}
    \label{fig:collision_rate}
\end{figure}

\noindent
Since the Lyapunov exponent is defined per unit time, a change in the collision rate with~$\alpha$ could contribute to the observed $\lambda(\alpha)$ trend. Figure~\ref{fig:collision_rate} shows the mean collision rate (collisions per unit time) as a function of~$\alpha$ for the stadium and Sinai geometries. The collision rate varies by approximately 20--25\% over $\alpha \in [0,1]$, which is significant but substantially smaller than the 66--76\% change in the Lyapunov exponent. Dividing the ensemble-averaged Lyapunov exponent by the collision rate yields the per-collision exponent $\lambda_{\rm coll}(\alpha) = \lambda(\alpha)/\nu(\alpha)$, which decreases by $57\%$ (stadium) and $70\%$ (Sinai) over the full $\alpha$ range. The per-collision exponent exhibits the same qualitative structure as the per-time exponent---smooth monotonic decrease for the stadium, and fast drop followed by plateau for the Sinai---confirming that the $\lambda(\alpha)$ trends are intrinsic to the dynamics and not an artifact of collision-rate changes.\\

\subsection{Conserved quantity verification}
\label{app:conserved_quantity}
\begin{figure}[t]
    \centering
    \includegraphics[width=\columnwidth]{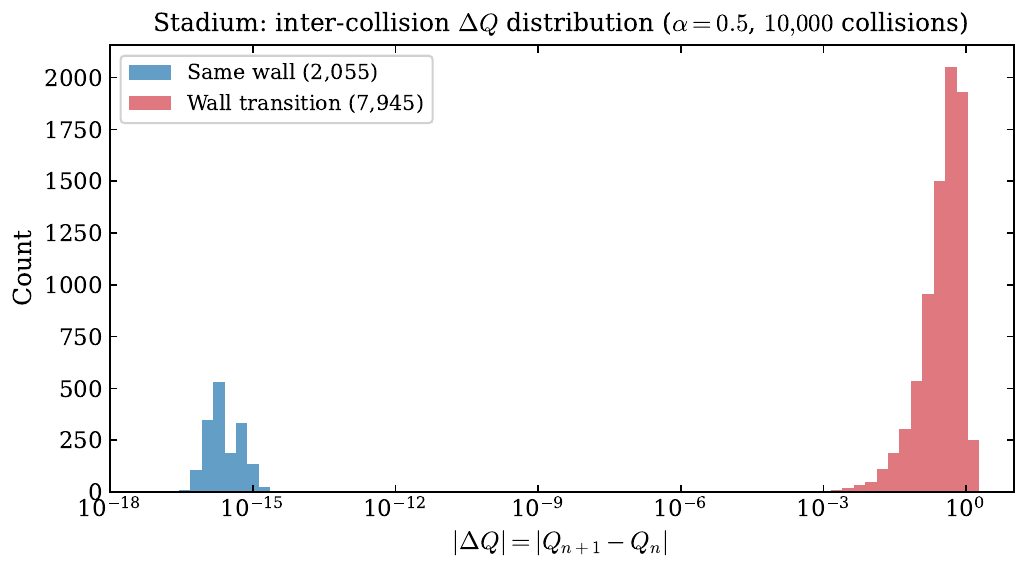}
    \caption{Distribution of inter-collision $|\Delta Q|$ in the stadium at $\alpha = 0.5$ ($10{,}000$ collisions). Same-wall consecutive pairs (blue) cluster at machine precision, while wall-transition pairs (red) show $\mathcal{O}(1)$ jumps---confirming the analytical prediction and the correctness of the collision-law implementation.}
    \label{fig:conserved}
\end{figure}

\noindent
To verify the per-collision conservation of $Q = v_\parallel - \alpha u$ (Eq.~\ref{eq:Q_conservation}), we track the inter-collision change $|\Delta Q| = |Q_{n+1} - Q_n|$ across $10{,}000$ collisions in the stadium at $\alpha = 0.5$. Figure~\ref{fig:conserved} shows the distribution of $|\Delta Q|$, classified by whether consecutive collisions hit the same wall or involve a wall transition. Same-wall pairs cluster at machine precision ($\sim 10^{-16}$), confirming exact conservation, while wall-transition pairs show $\mathcal{O}(1)$ jumps due to the change in tangent direction---a separation of ${\sim}\,16$ orders of magnitude.

\subsection{Datseris--Hupe--Fleischmann scaling test}
\label{app:dh_scaling}
\noindent
Datseris, Hupe, and Fleischmann~\cite{datseris2019} observed that in standard billiards, the mean Lyapunov exponent scales inversely with the chaotic phase-space volume fraction: $\lambda \propto 1/f_{\rm chaotic}$. Physically, this means that shrinking the chaotic sea concentrates the same total instability into fewer trajectories. We test whether this scaling extends to spinning billiards by computing both $\lambda(\alpha)$ (from the precomputed $30{,}000$-IC sweep) and $f_{\rm chaotic}(\alpha)$ (from FTLE distributions with a threshold of $0.01$).\\

\begin{figure}[t]
    \centering
    \includegraphics[width=\columnwidth]{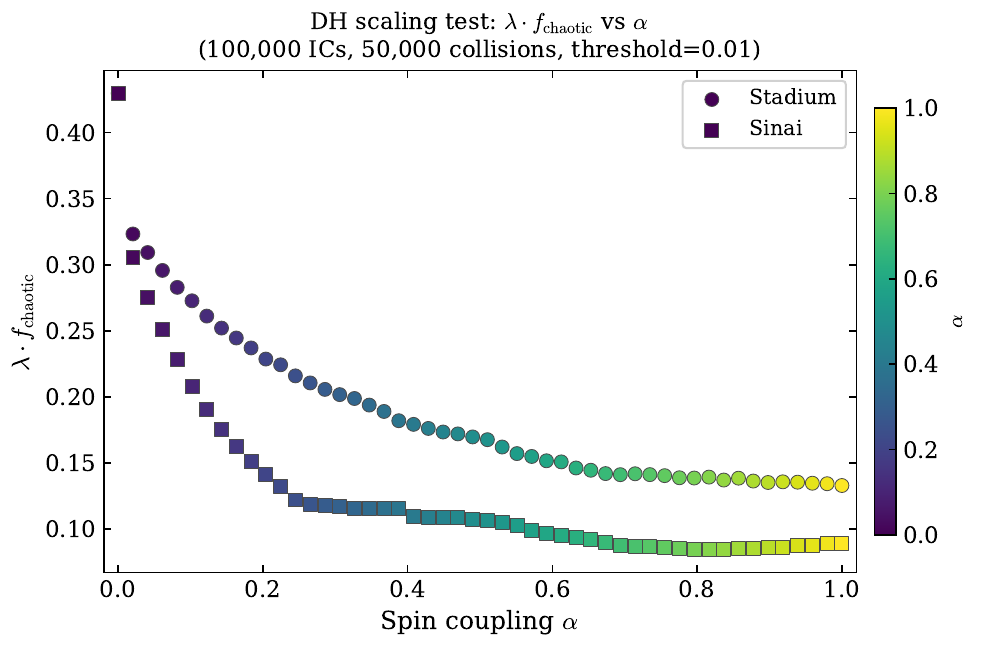}
    \caption{Test of the Datseris--Hupe--Fleischmann scaling $\lambda \propto 1/f_{\rm chaotic}$. The product $\lambda \cdot f_{\rm chaotic}$ vs~$\alpha$ ($10^5$ ICs, $5\times 10^4$ collisions, threshold $= 0.01$): DH scaling predicts this product to be constant, but it drops by a factor of~2--3 across the range of~$\alpha$. Circles: stadium; squares: Sinai; color indicates~$\alpha$ (viridis scale).}
    \label{fig:dh_scaling}
\end{figure}

\noindent
Figure~\ref{fig:dh_scaling} shows that the DH scaling clearly fails: the product $\lambda \cdot f_{\rm chaotic}$ decreases substantially with~$\alpha$, rather than remaining constant. The chaotic fraction drops only modestly (from $100\%$ to $\sim\!85\%$--$87\%$), while the Lyapunov exponent decreases by a factor of $3$--$4$. This demonstrates that spin coupling reduces the \emph{intensity} of chaos across the chaotic subset---not merely the fraction of chaotic trajectories---because the conserved quantity~$Q$ acts as a partial constraint that slows divergence rates even for orbits that never become fully regular.

\subsection{Full Lyapunov spectrum}
\label{app:spectrum}
\noindent
The spinning billiard has a 5D state space $(x,y,v_x,v_y,u)$ with energy conservation, yielding four non-trivial Lyapunov exponents on the energy shell plus one strongly negative exponent corresponding to off-shell perturbations. For a time-reversible conservative system, the spectrum must satisfy the pairing rule $(\lambda_1, \lambda_2, 0, -\lambda_2, -\lambda_1)$.\\

\begin{figure}[t]
    \centering
    \includegraphics[width=\columnwidth]{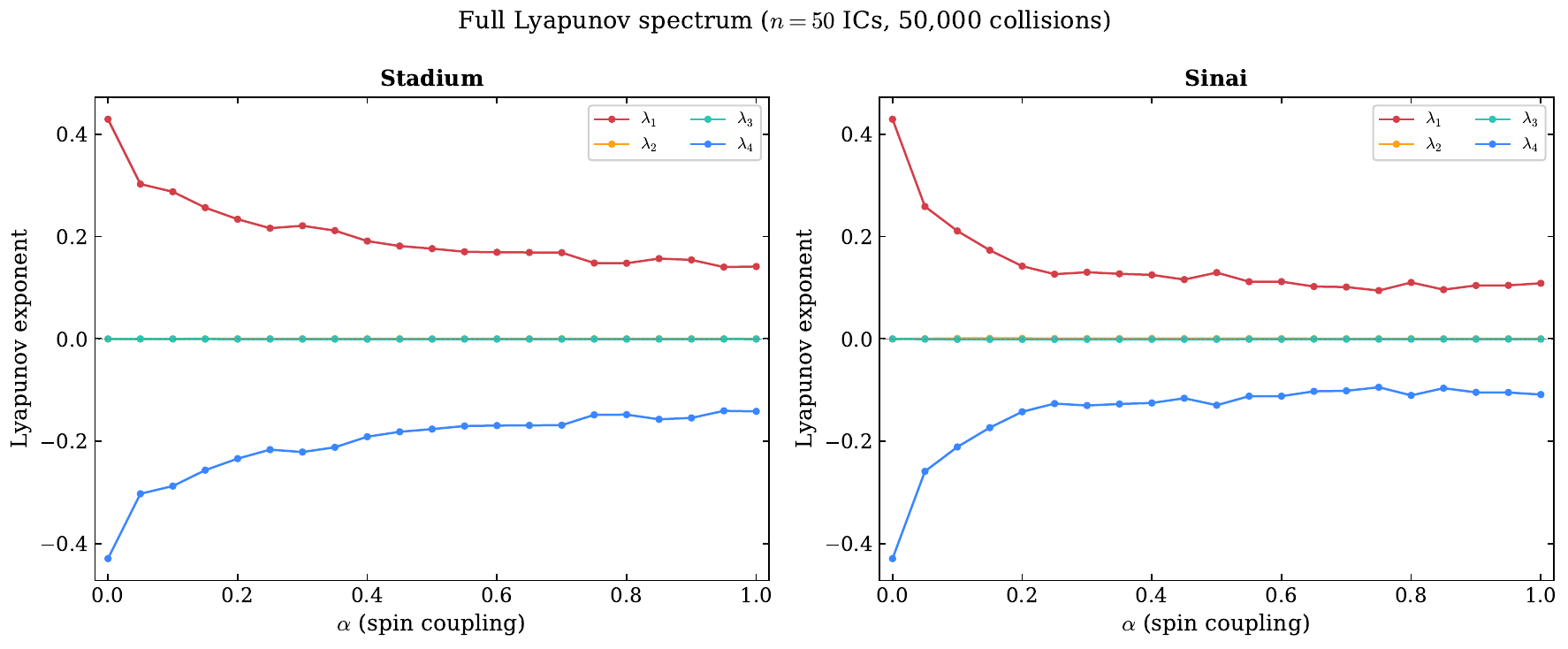}
    \caption{Full Lyapunov spectrum (4 on-shell exponents) vs~$\alpha$ for the stadium and Sinai geometries (50 ICs, $5\times 10^4$ collisions). The spectrum satisfies the pairing rule $\lambda_1 \approx -\lambda_4$, $\lambda_2 \approx -\lambda_3 \approx 0$ at all~$\alpha$. The absence of a second positive exponent ($\lambda_2 \approx 0$) confirms that spin does not create hyperchaos.}
    \label{fig:spectrum}
\end{figure}

\noindent
We compute the full spectrum using the Benettin algorithm with modified Gram-Schmidt orthogonalization, tracking 5 perturbation vectors simultaneously ($50$ ICs, $5\times 10^4$ collisions). Figure~\ref{fig:spectrum} shows the four on-shell exponents as a function of~$\alpha$. The pairing $\lambda_1 \approx -\lambda_4$ and $\lambda_2 \approx -\lambda_3 \approx 0$ is satisfied to high accuracy at all~$\alpha$, confirming both the time-reversal symmetry and the energy-shell structure.\\

\noindent
Crucially, $\lambda_2 \approx 0$ for all~$\alpha$, indicating that spin coupling does \emph{not} create hyperchaos: the dynamics has only a single expanding direction, just as in the standard spinless billiard. At $\alpha=0$ this is expected (spin is decoupled, so the spin direction is neutral), but it persists at $\alpha > 0$ where spin and translation are fully coupled. The $Q$ conservation mechanism provides a natural explanation: on same-tangent collision sequences, the additional constraint reduces the effective dynamics to 2D, precluding a second expanding direction.\\

\noindent
We note that the Kaplan--Yorke dimension is not a useful diagnostic in this setting: for a conservative system that preserves phase-space volume, the Lyapunov exponents sum to zero by construction, and the Kaplan--Yorke dimension equals the phase-space dimension regardless of whether the dynamics is chaotic or regular. The informative quantity is instead the number of positive exponents and their magnitudes, as reported above.

\end{document}